# Positioning Information Based Technique in Cooperative MIMO-OFDM Systems

Y. Nasser *member IEEE*, H. Farhat *member IEEE*, J.-F. Hélard *Senior member IEEE*

Institute of Electronics and Telecommunications of Rennes, UMR CNRS 6164, Rennes, France

Email : youssef.nasser@insa-rennes.fr

**ABSTRACT—** *Future Communication networks are tending towards a diverse wireless networking world where the positioning information (PI) could be helpful in different techniques like the dynamic resource allocation. On the other hand, the PI could be widely used for cooperative techniques in the relay and/or routing selection process. In this paper, we propose to use the PI in the selection of the relays and then to apply an efficient double layer distributed space time block code (DLSTBC) scheme between the different relays. Using the amplify and forward (AF) technique, we show that the proposed code is very efficient whatever the transmitted power is. Moreover, we show that the relay selection process based on PI yields very powerful results when compared to the random relay selection (RS) process.*

## 1. INTRODUCTION

Future communication networks are tending towards a diverse wireless networking world, where scenarios define that the user will be able to attain any service, at any time on effectively any network that is optimized for the application at hand. In actual research works, one of the key parameters in the optimization strategies consists in considering the positioning information (PI) together with a database, such as used in finger printing techniques, to reduce the feedback overhead. Indeed, the actual transmission techniques consume a valuable bandwidth in the feedback from the mobile terminal (MT) to the base station (BS). Moreover, the localization information could be used efficiently in the dynamic resource allocation, in synchronization and for interference cancellation purposes.

In cooperative networks, the PI could be widely used in the relays and routing selection. Also, PI and tracking mechanisms allow for predictions of future geographic node positions, hence help in deciding whether to use an advanced communication technique. Moreover, it could be widely used in some critical MT situations like tunnels, canyon regions as well as indoor areas.

Recently, the cooperative techniques have taken a lot of attention due to their efficiency in different transmission scenarios. The first research results on the relaying networks were obtained in the seventies in [1][2][3] and their interest was re-launched at the end of nineties [4][5][6] with the increase of services demands. It has then launched and activated a large amount of work in this domain. The cooperative techniques have widely used in different domains and their combination with other efficient techniques could be of great interest in the future wireless networks. Among these techniques, the combination of multiple input multiple output (MIMO) scheme with cooperative techniques, called hereafter COOPMIMO, has obtained a lot of attention. Contrarily to the classical techniques, the COOPMIMO system is based on the use of relays between the BS and the MT in such a way that the MIMO scheme is applied in a distributed manner among the different relays. In the literature, some works have been achieved in this area [9][10][11]. In [9] and [10], it has been proposed a general description of the COOPMIMO techniques with decode-and-forward [7] (DF), amplify-and-forward [8] (AF) algorithms and some analytical computations were given. However, the results with both algorithms were achieved without channel coding. The authors of [11] show how cooperation among nodes of a wireless network can be useful to reduce the overall radiated power necessary to guarantee reliable links among the network nodes. However, to the best knowledge of the authors, no contribution has been presented on COOPMIMO using PI. The reader may also refer to [12][13] for other contributions.

The work presented in this paper has been carried out within the framework of the European FP7-IST-WHERE project. The contribution of this work is multifold. First, we propose PI based technique to select the relays among the competitive relays existing in the network. Cooperating nodes can be specifically assigned fixed, or portable relay nodes, or other user terminals that temporarily form an ad-hoc point-to-point link. Then, we propose to apply a distributed space-time (ST) encoding scheme between the selected relays to ensure the full COOPMIMO orthogonal frequency division multiplexing (OFDM) system and we propose a generalized transmission model from the BS to the MT using the AF technique. Therefore, we analyze and compare some of the most promising MIMO systems [14][15] found in the literature with a double layer ST (DLST) scheme proposed in our previous work [16]. The DLST is selected due to its efficiency face to power imbalance between the transmitting antennas. It will be applied in a distributed way between different relays.

This paper is structured as follows. Section 2 describes the PI based relay selection technique. In section 3 we present the COOPMIMO system model using the AF technique. In section 4, we present the receiver system model using iterative detector. The system performance is evaluated in section 4 where conclusions are drawn in section 5.

## 2. PI BASED RELAY SELECTION TECHNIQUE

The main problem in cooperative networks is to select the best relay node in such a way that the cooperative technique is efficient in complex situations like tunnels and canyons. Indeed, in real networks, when the MT is moving in urban or suburban areas through deep shadowing regions for a long time, the received signal power is really attenuated and then the connection would be lost. Therefore, the general layout of the movement areas as well as the PI could be of great help in ensuring the continuity of the signal power level and hence establishing the connection.

In this work, we consider a network of *N* fixed relays uniformly distributed in a cell of diameter *D* meters. We assume that the transmission from the BS to the MT is achieved through a two-hop relay link by using *R*=2 or *R*=4 relays among the *N* competitive relays. We assume that each relay has one transmit/receive antenna where the destination has *M*=2 receiving antennas. Moreover, we assume that the MT is moving through a trajectory of *K* meters along the x-axis containing some deep shadowing regions like tunnels and that the PI of the *N* relays as well as of the MT are available at the BS (see Figure 1). Using the PI, the BS will select (and update each *m* meters) the *R* necessary relays to perform the COOPMIMO technique by computing the respective distances between the *N* relays and the MT each *m* meters. The *R* selected relays are chosen according to:

$$r = \min_{R_i}(d_{R_i,MT})$$
$$= \min_{R_i}\left(\sqrt{d_{BS,R_i}^2 + d_{BS,MT}^2 - 2.d_{BS,MT}.d_{BS,R_i}\cos(\theta)}\right) \quad (1)$$

where $d_{BS,MT}$, $d_{BS,R_i}$ and $d_{R_i,MT}$ are respectively the distances BS-MT, BS-relays and relays-MT.

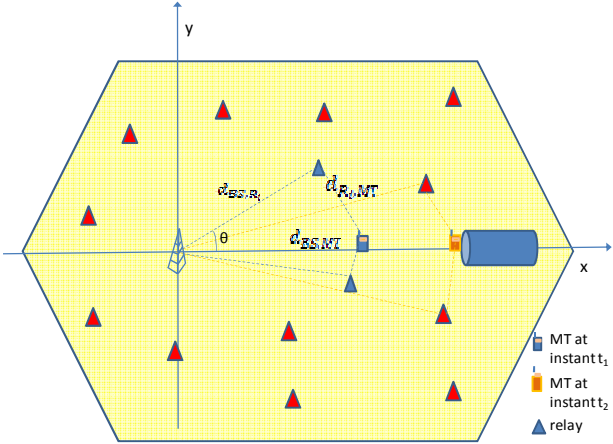

Figure 1- Real transmission scenario scheme

Once the choice of the cooperating relays is done, different techniques could then be applied in order to enhance the overall system performance in terms of bit rate, transmitted power, and connectivity. The problem turns out now to find an efficient COOPMIMO technique between the selected relays at each update distance. In the following, we will see how the number of competitive relays in the network as well as the cooperation technique will influence the results.

### 3. COOPMIMO SYSTEM MODEL

*3.1. Transmission model*

At the relay nodes, we assume that the AF technique is applied. In this technique, the relay amplifies and then retransmits to the destination the data it receives from the source. We assume that the transmission between different nodes is achieved by using the coded OFDM waveform with $N_s$ orthogonal sub-carriers. Figure 2 depicts the two-hop relaying scheme used in our study. At the source node, the information bits $b_k$ are first channel encoded using a convolutional encoder of rate $R_c$, randomly interleaved, and fed to a quadrature amplitude modulation (QAM) module. In this work, the transmission is modeled in frequency domain where $X[n]$ is the complex transmitted symbol on the $n^{th}$ sub-carrier obtained at the output of the QAM mapper. We assume in this paper that the source symbols are normalized i.e. $E\{|X[n]|^2\} = 1$ where E denotes the expectation operator.

The cooperation is done in two phases as follows (see Figure 2). Once the relays are selected based on the PI technique, the symbols are broadcasted to the MT and the selected relays. In phase 2, the selected relays retransmit the data to the MT. The update of the selection process will be done every m meters. That is, in phase 1 and at the receiving nodes (relays and MT), the frequency-domain received complex symbols at each relay node *r* are given by

$$Y_r = \sqrt{P_r}H_{s,r}X + V_r \quad (2)$$

and the received complex symbols, in phase 1, at each receiving antenna of the MT are given by:

$$U_{s,d} = \sqrt{P_d}H_{s,d}X + V_{s,d} \quad (3)$$

where $P_d$ is the received power by the destination (MT), $H_{s,d}$ and $H_{s,r}$ are respectively the frequency channel coefficients between the source(BS) and the MT, and between the BS and the relay. $V_{s,d}$ and $V_r$ denote the additive white Gaussian noise (AWGN) at the MT and relay node respectively with variance $\sigma^2$. Taking into account the transmission path loss, the received powers depending on the position, expressed in dB, are given by:

$$P_r = 10log10(P_s) - 32.4 + 10\alpha\log(d_{BS,r})$$
$$P_d = 10log10(P_s) - (-32.4 + 10\alpha\log(d_{BS,MT})) \quad (4)$$

where $P_s$ is the transmitted power by the BS and $\alpha$ is the propagation constant which depends on the transmission environment.

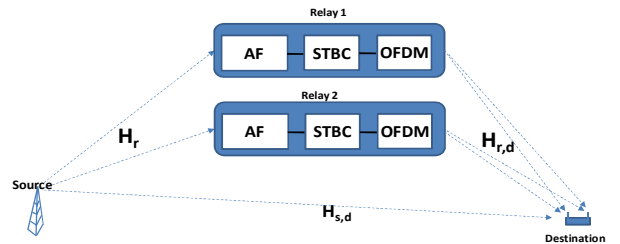

Figure 2- Two-hop COOPMIMO network

In phase 2, the *R* selected relays will retransmit the received data from the BS to the MT. Using the AF technique, we propose in this work to apply a distributed ST encoding scheme between the signals transmitted by the different relays. In our study, we limit the number of selected relays at each instant *t* to *R*=2 or 4 relays.

*3.1. Relay processing*

Using the AF technique, the selected relays normalize first the powers of the received data before applying the ST encoding scheme. The normalized received data obtained by (2) becomes:

$$Y_r = \frac{1}{\sqrt{P|H_r|^2 + \sigma^2}}(\sqrt{P}H_rX + V_r) \quad (5)$$

$$= H_r^{eq} X + H_r^{nor} V_r$$

Once the power normalized, each relay uses a ST encoder which takes $Q$ data complex symbols and transforms them to a ($R,T$) output matrix according to the ST block coding (STBC) scheme. The STBC coding rate is then defined by $L=Q/T$. In this process, each group $\mathbf{y_r} = [Y_{1,r}, ... Y_{Q,r}]$ (where $Y_{q,r}$ verifies (5)) of $Q$ complex symbols at each relay is fed to the ST encoder. Let $\mathbf{W_r} = [w_{i,t,r}]$ where $w_{i,t,r}$ ($i=1,...,R$; $t=1,...,T$) be the matrix corresponding to the output of the ST encoder at each relay $r$. In order to achieve the distributed MIMO transmission scheme, each relay $r$ selects the $r^{th}$ vector of the encoding matrix $\mathbf{W_r}$ and transmits it to the destination. The set of the $R$ vectors transmitted by all the relays will be modeled by the matrix $\mathbf{W}$ having the same dimensions of $\mathbf{W_r}$, i.e. ($R,T$). In this model, the COOPMIMO system could be seen as a conventional MIMO scheme with the encoding matrix $\mathbf{W}$. Moreover, in order to have a fair analysis between the different ST coding schemes, the output power of the ST encoder is normalized by $R$.

Let us now describe the transmission model between the relays and the destination independently of the ST scheme. Let $\mathbf{x}=[X_1,...,X_Q]$ of $Q$ complex symbols be the vector of the symbols transmitted by the source used to generate the vector $\mathbf{y_r} = [Y_{1,r}, ... Y_{q,r}, ..., Y_{Q,r}]$ at the input of the ST encoder at the different relays. The relationship between $\mathbf{x}$ and $\mathbf{y_r}$ can be deduced from (2) by:

$$\mathbf{y_r} = \mathbf{PH_r x} + \mathbf{v_r} \quad r=1,...,R \quad (6)$$

where $\mathbf{H_r}$ is a ($Q,Q$) diagonal matrix composed of the channel coefficients between the source and each relay $r$ and $\mathbf{P}$ is the ($Q,Q$) matrix of the powers transmitted by the source (assumed identical for all symbols).

By introducing an equivalent receive matrix at the second phase $\mathbf{Y_d} \in \mathbb{C}^{M \times T}$ whose elements are the complex received symbols from the relays, we can write the received signal on the receiving antennas at the destination as:

$$\mathbf{Y_d} = \mathbf{H_{r,d} P_{r,d} W} + \mathbf{V_d} \quad (7)$$

where $\mathbf{H_{r,d}}$ is the ($M,R$) channel matrix whose components are the coefficients $H_{j,r}$ between the relays and the destination, $\mathbf{W}$ is a ($R,T$) complex matrix containing the transmitted symbols $w_{i,t,r}$. $\mathbf{P_{r,d}}$ is the ($R,R$) matrix of the powers received by the MT from the different relays and $\mathbf{V_d}$ is a ($M,T$) complex matrix corresponding to the AWGN.

We separate the real and imaginary parts of the entries $\mathbf{y_r}$ in (6), of the outputs $\mathbf{W_r}$ of the ST encoder (of each relay $r$) as well as those of the channel matrix $\mathbf{H_{r,d}}$ and the received signal $\mathbf{Y_d}$. Let $Y_{q,r,\Re}$ and $Y_{q,r,\Im}$ be the real and imaginary parts of $Y_{q,r}$. The main parameters of the ST code are given by its dispersion matrices $\mathbf{U_q}$ and $\mathbf{V_q}$ ($q=1,...,Q$), corresponding (not equal) respectively to the real and imaginary parts of $\mathbf{W_r}$. With these notations, $\mathbf{W_r}$ is given by:

$$\mathbf{W_r} = \sum_{q=1}^{Q-1} Y_{q,r,\Re} \mathbf{U_q} + Y_{q,r,\Im} \mathbf{V_q} \quad (8)$$

We separate the real and imaginary parts of, $\mathbf{y_r}$, $\mathbf{Y_d}$ and $\mathbf{W}$ and stack them row-wise in vectors of dimensions (2$Q$,1), (2$Q$,1), (2$MT$,1) and (2$RT$,1) respectively. We obtain:

$$\begin{aligned}
\mathbf{x} &= [X_{1,\Re}, X_{1,\Im}, ..., X_{Q,\Re}, X_{Q,\Im}]^T \\
\mathbf{y_r} &= [Y_{1,r,\Re}, Y_{1,r,\Im}, ..., Y_{Q,r,\Re}, Y_{Q,r,\Im}]^T \\
\mathbf{y_d} &= [Y_{1,\Re}, Y_{1,\Im}, ..., Y_{T,\Re}, Y_{T,\Im}, ..Y_{MT,\Re}, Y_{MT,\Im}]^T \\
\mathbf{w} &= [w_{(1,1),\Re}, w_{(1,1),\Im}, ..., w_{(2R,T),\Re}, w_{(2RT,1),\Im}]^T
\end{aligned} \quad (9)$$

where $\mathbf{T}$ holds for matrix transpose.

Since, we use linear ST coding, vector $\mathbf{w}$ can be written as:

$$\mathbf{w} = \mathbf{F}.\mathbf{y} \quad (10)$$

where $\mathbf{y} = [\mathbf{y_1}, ..., \mathbf{y_r}, ..., \mathbf{y_R}]^T$ and $\mathbf{F}$ has the dimensions (2$RT$, 2$RQ$) and is obtained through the dispersion matrices of the real and imaginary parts of $\mathbf{s}$.

The problem now is to achieve the relationship between the transmitted vector by the source, i.e. $\mathbf{x}$, and the received vector by the destination $\mathbf{y_d}$. As we change the formulation of $\mathbf{x}$, $\mathbf{y_r}$, $\mathbf{Y_d}$ and $\mathbf{w}$ in (9), it can be shown that vectors $\mathbf{x}$ and $\mathbf{y_d}$ are related through the matrix $\mathbf{G}$ and $\mathbf{H^{eq}}$ of dimensions (2$MT$, 2$RT$) and (2$RT$, 2$Q$) such that:

$$\begin{aligned}
\mathbf{y_d} &= \mathbf{GBFH^{eq}x} + \mathbf{GFH^{nor}V_r} + \mathbf{V_d} \\
&= \mathbf{G_{eq}x} + \mathbf{V}
\end{aligned} \quad (11)$$

where matrix $\mathbf{G}$ is composed of blocks $\mathbf{G_{j,r}}$ ($j=1,...,M$; $r=1,...,R$) each having (2$T$,2$T$) elements given by:

$$\mathbf{G_{j,r}} = \begin{pmatrix} H_{j,r,\Re} & -H_{j,r,\Im} & 0 & 0 & \cdots & & 0 \\ H_{j,r,\Im} & H_{j,r,\Re} & 0 & 0 & & & \\ 0 & 0 & H_{j,r,\Re} & -H_{j,r,\Im} & 0 & \cdots & \vdots \\ 0 & 0 & H_{j,r,\Im} & H_{j,r,\Re} & 0 & \cdots & \\ \vdots & & & 0 & \ddots & 0 & 0 \\ \vdots & & & & \ddots & \ddots & 0 & 0 \\ & & & & & 0 & H_{j,r,\Re} & -H_{j,r,\Im} \\ 0 & \cdots & & & 0 & H_{j,r,\Im} & H_{j,r,\Re} \end{pmatrix}_{(2T,2T)} \quad (12)$$

where $H_{j,r,\Re}$ (respectively $H_{j,r,\Im}$) is the frequency channel coefficient between the $r^{th}$ relay and the $j^{th}$ receiving antenna. These channel coefficients are assumed to be invariant during $T$ OFDM symbols. $\mathbf{B}$ is a (2$RT$, 2$RT$) diagonal matrix corresponding to the powers transmitted by the relays. $\mathbf{G_{eq}}$ is the equivalent channel matrix, of dimensions (2$MT$, 2$Q$), between the transmitted data vector by the source $\mathbf{x}$ and the received vector by the destination through the relays. $\mathbf{H^{eq}}$ and $\mathbf{H^{nor}}$ of dimensions (2$RQ$, 2$RQ$) can be written in the same way of $\mathbf{G}$. They are easily obtained through (5).

## 4. RECEIVER MODEL

### 4.1. Receiver processing: ST decoder

The detection problem at the second phase is to find the transmitted data $\mathbf{x}$ given the vector $\mathbf{y_d}$ and to combine the estimated data at the second phase with that estimated data at the phase 1 in such a way to maximise the combining ratio. In our work, we combine the estimated values $\hat{X}_{q,1}$ and $\hat{X}_{q,2}$ as follows:

$$\hat{X}_q = \frac{\rho_1 \hat{X}_{q,1} + \rho_2 \hat{X}_{q,2}}{\rho_1 + \rho_2} \quad (13)$$

where $\rho_1$ and $\rho_2$ are the respective estimated SNR at each phase and $\hat{X}_{q,1}$ (resp. $\hat{X}_{q,2}$) is the estimated data at the first phase (resp. second phase). The question is now how to estimate the data at the second phase.

In order to estimate the received symbols in phase 2 and in the case of orthogonal STBC (OSTBC), the optimal receiver is made of a concatenation of ST decoder and channel decoder modules. In non-orthogonal STBC (NO-STBC) schemes, a sub-optimal solution composed of an iterative detector is proposed here as shown in Figure 3.

At the first iteration, the demapper takes the estimated symbols $\hat{\mathbf{s}}$, the knowledge of the channel $\mathbf{G_{eq}}$ and of the noise variance, and computes the log likelihood ratio (LLR) values of each of the coded bits transmitted per channel use. The estimated symbols $\hat{\mathbf{s}}$ are obtained via minimum mean square error (MMSE) filtering according to:

$$\hat{X}_{p,2}^{(1)} = \mathbf{g_p^T}\left(\mathbf{G_{eq}}.\mathbf{G_{eq}^T} + \sigma^2\mathbf{I}\right)^{-1}\mathbf{y_d} \quad (14)$$

where $\mathbf{g_p^T}$ of dimension ($2MT$, 1) is the $p^{th}$ column of $\mathbf{G_{eq}}$ ($1 \leq p \leq 2Q$). $\hat{X}_{p,2}^{(1)}$ is the estimation of the real part ($p$ odd) or imaginary part ($p$ even) of $X_q$ ($1 \leq q \leq Q$) at the second phase. This estimation will be combined with the estimation of the transmitted symbol at the first phase according to (13). Once the estimation of the different symbols $X_q$ is achieved by the soft mapper at the first iteration, we use this estimation for the next iterations process.

From the second iteration, we perform parallel interference cancellation (PIC) operation followed by a simple inverse filtering:

$$\hat{\mathbf{y}}_{\mathbf{p}} = \mathbf{y_d} - \mathbf{G_{eq,p}}\tilde{\mathbf{X}}_p^{(\ell-1)}$$
$$\tilde{\mathbf{X}}_p^{(\ell)} = \frac{1}{\mathbf{g_p^T}\mathbf{g_p}}\mathbf{g_p^T}\hat{\mathbf{y}}_{\mathbf{p}} \quad (15)$$

where $\mathbf{G_{eq,p}}$ of dimension ($2MT$, $2Q$-1) is the matrix $\mathbf{G_{eq}}$ with its $p^{th}$ column removed, $\tilde{\mathbf{X}}_{p,2}^{(\ell-1)}$ of dimension ($2Q$-1, 1) is the vector $\tilde{X}$ estimated by the soft mapper at the previous iteration with its $p^{th}$ entry removed.

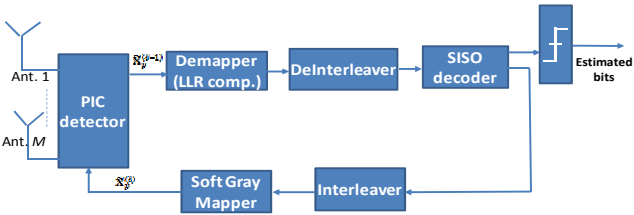

Figure 3- Iterative receiver structure

### 4.2. Considered ST coding schemes

The aim of this section is to judiciously build the ST code so that the resulting MIMO scheme behaves efficiently in different transmission scenarios. In [16], we have proposed a DLST code for broadcasting systems which is robust in unequal received power scenarios. We then exploit this conclusion to study the DLST in the COOPMIMO systems. In the sequel, we will first consider the well-known orthogonal Alamouti ST coding scheme [14] for its robustness and its simplicity. For NO schemes and $R$=2, we consider the full rate and the fully diverse Golden code [15].

For $R$= 4 relays, we implement the DLST proposed in [16]. This DLST is robust against unbalanced received power and is constructed by using the Alamouti code at the first layer and the Golden code at the second layer. It is given by:

$$\mathbf{W} = \frac{1}{\sqrt{5}}\begin{pmatrix} \alpha(s_1+\theta s_2) & \alpha(s_3+\theta s_4) & \alpha(s_5+\theta s_6) & \alpha(s_7+\theta s_8) \\ j\bar{\alpha}(s_3+\bar{\theta}s_4) & \bar{\alpha}(s_1+\bar{\theta}s_2) & j\bar{\alpha}(s_7+\bar{\theta}s_8) & \bar{\alpha}(s_5+\bar{\theta}s_6) \\ -\alpha^*(s_5^*+\theta^*s_6^*) & -\alpha^*(s_7^*+\theta^*s_8^*) & \alpha^*(s_1^*+\theta^*s_2^*) & \alpha^*(s_3^*+\theta^*s_4^*) \\ j\bar{\alpha}^*(s_7^*+\bar{\theta}^*s_8^*) & -\bar{\alpha}^*(s_5^*+\bar{\theta}^*s_6^*) & -j\bar{\alpha}^*(s_3^*+\bar{\theta}^*s_4^*) & \bar{\alpha}^*(s_1^*+\bar{\theta}^*s_2^*) \end{pmatrix} \quad (16)$$

where $\theta = \frac{1+\sqrt{5}}{2}, \bar{\theta} = 1-\theta, \alpha = 1+j(1-\theta), \bar{\alpha} = 1+j(1-\bar{\theta})$.

## 5. SYSTEM PERFORMANCE

In this section, we validate the proposed technique by simulation results. The simulation parameters considered in this paper are summarized in Table 1. In the simulations obtained hereafter, we consider that the links between the different nodes are achieved through a Rayleigh i.i.d frequency channel model. The cooperation is achieved with AF technique and various transmitted powers. The results are obtained with a spectral efficiency $\eta$= 4 b/s/Hz. These values are obtained as given in Table 2. We assume that the MT is traveling through a trajectory of 1 Km having a tunnel of 200 m. The propagation loss in the tunnel is characterized by two propagation regions as de described in [17]. The first region of 50 meters has a power loss of 15 dB. The second one is generally modeled as a wave guide having a power degradation of roughly 0.06 dB/meter. During its movement, we assume that the MT updates the selected relays according to (1) each 10 m.

Table 1- Simulations Parameters

| Simulation parameter | Value |
|---|---|
| Bandwidth ($f_s$=1/$T_s$) | 20 MHz |
| Rate $R_c$ of conv. code | 1/2, 2/3 |
| Polynomial code generator | (133,171)$_o$ |
| Channel estimation | perfect |
| Constellation | QPSK, 16-QAM, 64-QAM |

Table 2- Different MIMO schemes and efficiencies

| Spectral Efficiency | ST scheme | ST rate L | Constellation | R |
|---|---|---|---|---|
| $\eta$=4 b/s/Hz | Alamouti | 1 | 64-QAM | 2/3 |
| | Golden | 2 | 16-QAM | 1/2 |
| | DLST | 2 | 16-QAM | 1/2 |

Figure 4 and Figure 5 give the measured bit error rate (BER) at the destination with respect to the number of competitive relays existing in the network using respectively two values of transmitted powers $P_s$= 0dB and $P_s$= 2dB. Figure 4 shows that the PI based technique gives better performance than the random relay selection (RS) based technique whatever the number of competitive relays is. Moreover, the last conclusion could be verified indirectly as well. Indeed, when the number of competitive relays increases, the MT could choose at each update time the closest $R$ relay nodes and hence allows an increased received power. The system converges when the number of available relay nodes in the network is greater than 100.

Eventually, we show that the proposed code DLST is very efficient when compared with Alamouti and Golden codes since it is designed to be robust face to unequal received powers [16]. Figure 5 gives the same conclusions as in Figure 4 and shows the efficiency of our techniques.

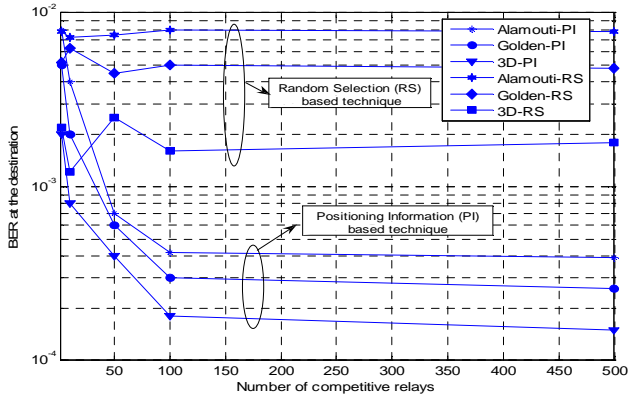

Figure 4- BER measured at the destination, transmitted power $P_s$= 0dB, η= 4 b/s/Hz

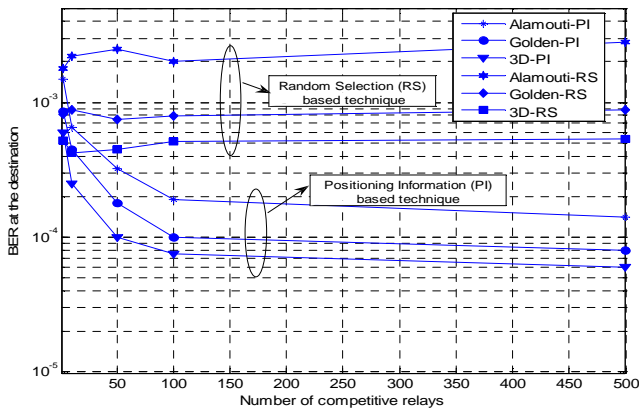

Figure 5- BER measured at the destination, transmitted power $P_s$= 2dB, η= 4 b/s/Hz

## 6. CONCLUSION

In this paper, we have proposed a generalized framework of the COOPMIMO system based on PI using the AF technique. We have shown that the PI information is powerful in the relay selection process. We have also proposed a very efficient double layer space time code to apply in a distributed way among the relays. This performance is verified in different scenarios whatever the transmitted power is. The DLST code is then very promising candidate for cooperation in OFDM systems. Moreover, its BER value decreases when it is used with PI based technique.

### ACKNOWLEDGMENTS

This work has been performed in the framework of the ICT project ICT-217033 WHERE, which is partly funded by the European Union.